\documentclass[a4paper,12pt]{article}
\usepackage{epsfig}
\usepackage{graphicx}
\newcommand {\kgdt}  {15.1}
\newcommand {\kgdc}  {8.6}
\newcommand {\kgdeff} {7.4}

\newcommand {\numcin}{9.8}

\newcommand {\proscin}{0.006}
\newcommand {\numqua}{6.2}
\newcommand {\probqua}{99.8}
\newcommand {\prosqua}{0.2}
\begin{document}
\begin{titlepage}

\begin{flushright}
       LYCEN 2002-31  \\
       June 14th, 2002
\end{flushright}

\vfill

{\bf\LARGE
\begin{center}
           Improved Exclusion Limits from the EDELWEISS
           WIMP Search
\end{center}}

\vfill

\begin{center}
{\large The EDELWEISS Collaboration:} \\
  A.~Benoit$^{1}$,
  L.~Berg\'e$^{2}$,
  A.~Broniatowski$^{2}$,
  L.~Chabert$^{3}$,
  B.~Chambon$^{3}$,
  M.~Chapellier$^{4}$,
  G.~Chardin$^{5}$,
  P. Charvin$^{5,6}$,
  M.~De~J\'esus$^{3}$,
  P. Di~Stefano$^{3}$,
  D.~Drain$^{3}$,
  L.~Dumoulin$^{2}$,
  J.~Gascon$^{3}$,
  G.~Gerbier$^{5}$,
  E.~Gerlic$^{3}$,
  C.~Goldbach$^{7}$,
  M.~Goyot$^{3}$,
  M.~Gros$^{5}$,
  J.P.~Hadjout$^{3}$,
  S.~Herv\'e$^{5}$,
  A.~Juillard$^{2}$,
  A.~de~Lesquen$^{5}$,
  M.~Loidl$^{5}$,
  J.~Mallet$^{5}$,
  S.~Marnieros$^{2}$,
  O.~Martineau$^{3}$,
  N.~Mirabolfathi$^{2}$,
  L.~Mosca$^{5,6}$,
  X.-F.~Navick$^{5}$,
  G.~Nollez$^{7}$,
  P.~Pari$^{4}$,
  C.~Riccio$^{5,6}$,
  V.~Sanglard$^{3}$,
  M.~Stern$^{3}$,
  L.~Vagneron$^{3}$
\end{center}

{\scriptsize\noindent
$^{1}$Centre de Recherche sur les Tr\`es Basses Temp\'eratures,
      SPM-CNRS, BP 166, 38042 Grenoble, France\\
$^{2}$Centre de Spectroscopie Nucl\'eaire et de Spectroscopie de Masse,
      IN2P3-CNRS, Universit\'e Paris XI,
      bat 108, 91405 Orsay, France\\
$^{3}$Institut de Physique Nucl\'eaire de Lyon-UCBL, IN2P3-CNRS,
      4 rue Enrico Fermi, 69622 Villeurbanne Cedex, France\\
$^{4}$CEA, Centre d'\'Etudes Nucl\'eaires de Saclay,
      DSM/DRECAM, 91191 Gif-sur-Yvette Cedex, France\\
$^{5}$CEA, Centre d'\'Etudes Nucl\'eaires de Saclay,
      DSM/DAPNIA, 91191 Gif-sur-Yvette Cedex, France\\
$^{6}$Laboratoire Souterrain de Modane, CEA-CNRS, 90 rue Polset,
      73500 Modane, France\\
$^{7}$Institut d'Astrophysique de Paris, INSU-CNRS,
      98 bis Bd Arago, 75014 Paris, France
}

\vfill

\begin{center}{\large\bf Abstract}\end{center}

The EDELWEISS experiment has improved its sensitivity
for the direct search for WIMP dark matter.
In the recoil energy range relevant for
WIMP masses below 10 TeV/c$^2$,
no nuclear recoils 
were observed in the fiducial volume
of a heat-and-ionization cryogenic Ge detector operated in
the low-background environment of
the Laboratoire Souterrain de Modane
in the Fr\'ejus Tunnel,
during an effective exposure of \kgdeff\ kg$\cdot$d.
This result is combined with the previous EDELWEISS data
to derive a limit on the cross-section for
spin-independent interaction of WIMPs and nucleons
as a function of WIMP mass,
using standard nuclear physics and astrophysical assumptions.
This limit excludes at more than \probqua\%CL
a WIMP candidate with a mass of 44 GeV/c$^2$ and a cross-section
of 5.4$\times 10^{-6}$~pb,
as reported by the DAMA collaboration.
A first sample of supersymmetric models are also excluded
at 90\%CL.

\vfill

%\begin{center}
% {\Large Submitted to Phys. Lett. B}
%PACS:
%95.35.+d, %Dark matter
%14.80.Ly, %Supersymmetric partners of known particles,
%98.80.Es, %Observational cosmology,
%29.40.Wk. %Solid-state detectors.
%Keywords: Dark Matter, WIMPs, Supersymmetry,
%Germanium Detectors, Bolometers.
%\end{center}

\end{titlepage}

%%%%%%%%%%%%%%%%%%%%%%%%%%%%%%%%%%%%%%%%

% \section{Introduction}
\noindent{\large\bf Introduction} \\

The experimental efforts in the search for Cold Dark Matter
in the form of Weakly Interacting Massive Particles (WIMPs)
are steadily increasing (see e.g. Ref.~\cite{bib-review}
for a review).
In direct searches, the experimental signature of
the WIMPs from the galactic halo is the observation of
nuclear recoils induced by their scattering.
Current experimental sensitivities for the interaction rate
of WIMPs are of the order of 1 per kilogram detector material
and per day for the various experiments at the forefront of this
%search~\cite{bib-nai,bib-dama,bib-hdms,bib-igex,bib-cdms,bib-cdms2,bib-edw2000}.
search~\cite{bib-nai}-\cite{bib-edw2000}.

The experiment DAMA~\cite{bib-dama}
has reported an annual modulation signal in NaI detectors.
This represents a challenge to other detecting methods
to reach equivalent sensitivities,
a standard procedure to compare different
experiments having been laid out in
Ref.~\cite{bib-lewin}.
Two experiments\cite{bib-cdms,bib-edw2000},
both using cryogenic heat-and-ionization germanium detectors,
were able to exclude at more than 90\%CL
the central value deduced by DAMA from
its annual modulation signal
for the WIMP mass and its nucleon scattering cross-section
($M_W$ = 52~GeV/c$^2$ and
$\sigma_n$ = $7.2\times10^{-6}$~pb, respectively).
The CDMS experiment~\cite{bib-cdms} was the first to report
a limit excluding this value.
However, the operation of the detectors in
a shallow site, with only 16 meters of water equivalent (m.w.e.)
protection from cosmic rays, leads to a sizable
background of nuclear recoils from neutron scattering
that requires a delicate procedure of
background subtraction~\cite{bib-cdms2}.
The EDELWEISS experiment\cite{bib-edw2000},
located in a 4800 m.w.e. deep underground site,
was also able to reject that value
without requiring any background subtraction.
However, the energy resolution in that experiment
restricted the sensitivity to nuclear recoils
above 30 keV and
the accumulated run time
was not sufficient to extend the sensitivity to
the central value obtained by DAMA~\cite{bib-dama}
when their annual modulation signal is combined
with their exclusion
limit from pulse shape discrimination in NaI~\cite{bib-nai}
($M_W$ = 44~GeV/c$^2$ and
$\sigma_n$ = $5.4\times10^{-6}$~pb).
Beyond this, an important step in these searches
would be to reach out to
the Supersymmetric model calculations
predicting the largest $\sigma_n$ values.

Following the results obtained by EDELWEISS using a
heat-and-ionization cryogenic Ge detector\cite{bib-edw2000},
three new detectors were put in operation.
The aim was to improve our understanding of
the performance of such detectors,
and to extend the sensitivity to lower
cross-sections.
This letter presents the improved cross-section limit
achieved using a detector with improved
charge collection and energy resolution.

\mbox{}\\\noindent{\large\bf Experimental Setup} \\
% Experimental setup

The experiment is located in the Laboratoire Souterrain de Modane
in the Fr\'ejus Tunnel under the French-Italian Alps,
under a 4800 m.w.e. rock overburden.
The experimental setup is described in~\cite{bib-edw2000};
only the relevant modifications are discussed here.

Three new 320~g cryogenic Ge
detectors~\cite{bib-navick},
each 70~mm in diameter and 20~mm in height,
are operated simultaneously.
Each one is equipped with a segmented electrode
defining two regions, a central part and a guard ring.
To improve their reliability, all electrical contacts
with the electrodes are ultrasound-bonded
instead of glued as in Ref.~\cite{bib-edw2000}.

Two of the detectors, labeled GeAl9 and GeAl10,
are very similar to the
one used in the year 2000 runs (Ref.~\cite{bib-edw2000}),
labeled GeAl6.
The third detector, GGA1, differs by the presence
of a 60~nm hydrogenated amorphous Ge layer
deposited under the 70~nm Al electrodes
and on all exposed surfaces.
This modification was done to test whether an amorphous
layer can improve charge collection properties,
as suggested by~\cite{bib-shutt}.
%As in Ref.~\cite{bib-edw2000}, the electrodes are regularly
%shorted to prevent charge accumulation
%due to trapping in the detector volume. 
Bias voltage values between 2 and 4~V are used.

The size of the NTD heat sensors and the thermalization
of the detectors are improved for a better sensitivity,
in light of the previous experience with GeAl6.
As a result, it was possible to operate the detector
at a reduced temperature of 17 mK
(regulated to within 10 $\mu$K).

The data acquisition system has been upgraded to
a design with fully numerical data flow and trigger.
The signals from the 3 heat and 6 ionization channels
are continuously digitized at respective rates of 2 
and  200 ksample/s
and sent to the data acquisition PC via an optical link.
The ionization data are then filtered on-line using an
Infinite Impulse Response (IIR) high-pass elliptic filter
of 4th order, in order to remove most of the microphonics noise,
below a frequency of 1200 Hz.
The trigger is defined by requiring a minimum threshold
on the absolute value of any of the filtered ionization
channels.

%%%%%%%%%%%%%%%%%%%%%%%%%%%%%%%%%%%%%%%%%%%%%%%%%%%%%%%%%%

\mbox{}\\\noindent{\large\bf Detector Calibration}\\

\begin{figure}[tbp]
\epsfig{file=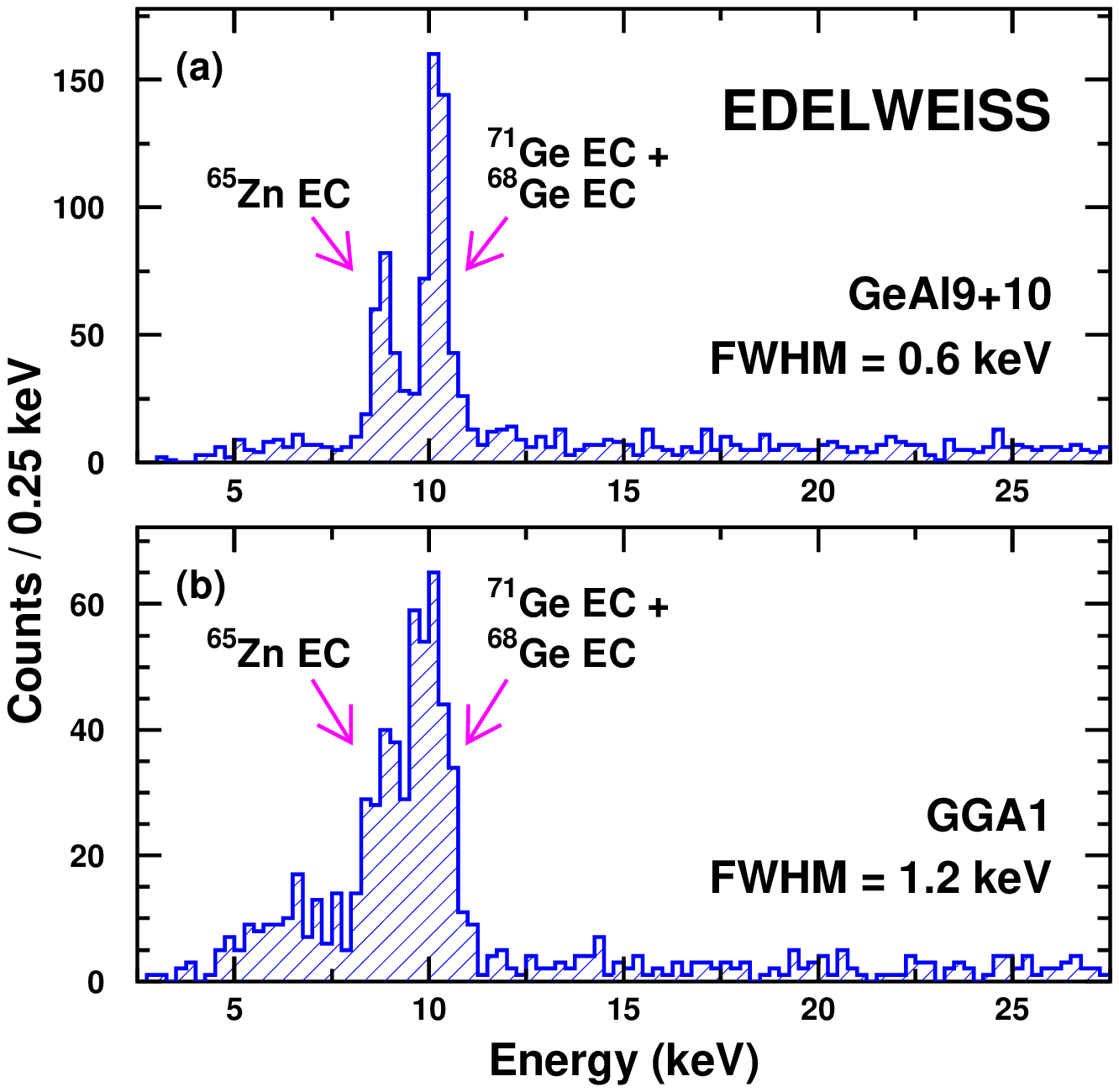
,height=17cm,bbllx=4cm,bblly=9cm,bburx=22cm,bbury=28cm}
\caption[]{Energy pulse height spectra for low-energy gammas
(sum of the ionization and heat channels, weighted by their
resolution squared) in the fiducial volume of the EDELWEISS detector,
for the low-background physics runs: (a) sum of the distributions
of the detectors GeAl9 and GeAl10; (b) distribution in
the detector GGA1.
The arrows indicate the peaks at 8.98 and 10.37 keV, corresponding
to the de-excitation of the cosmogenic activation of $^{65}$Zn
and $^{68}$Ge in the detector, and the $^{71}$Ge activation
that follows neutron calibrations.}
\label{fig-x} 
\end{figure}

The heat and ionization responses to $\gamma$ rays were
calibrated using $^{57}$Co and $^{60}$Co sources.
In 2000~\cite{bib-edw2000}, the performance of the
detector GeAl6 was partly limited by a poor baseline
resolution on both ionization and heat channels.
While resolutions at 122 keV of the new detectors
remained close to those of GeAl6,
the baseline resolutions were somewhat improved.
The ionization baseline resolutions are all below
1.5 keV FWHM,
and are 1.3, 0.5 and 0.4 keV for the heat channels in
GGA1, GeAl9 and GeAl10, respectively.
The resulting improvement is illustrated in
Fig.~\ref{fig-x}, showing the low-energy spectra
recorded in the three detectors in the low-background
physics run.
Here, the energy corresponds to the average of
the ionization and heat signals, weighted by the
square of their respective resolutions.
The 8.98 and 10.37 keV lines from the decay of
the cosmic-ray induced long-lived isotopes $^{65}$Zn
and $^{68}$Ge~\cite{bib-nuclide} are clearly resolved
in GeAl9 and GeAl10,
with a resolution of 0.6 keV FWHM.
The resolution in GGA1 is only 1.2 keV FWHM,
but the two-peaked structure can again be observed.
The degraded resolution of GGA1 relative to
the two other detectors is due to a reduced
NTD sensor volume (1.6 vs 5.6 mm$^3$), and
an increased sensitivity to microphonics
of the center electrode ionization channel. 

The threshold level of the ionization trigger was measured using
two different techniques.
The first one consists in extracting,
as in Ref.~\cite{bib-edw2000}, the threshold value corresponding
to an efficiency of 50\% from a fit to the low-energy part
of the Compton plateau recorded with a $^{60}$Co
$\gamma$-ray source.
The second technique was made possible by the simultaneous
operation of the three detectors with a $^{252}$Cf
neutron source.
Neutron scattering induces a large number of coincidence events
where at least two detectors are hit.
The efficiency curve as a function of ionization energy
in one detector is given by the ratio of the ionisation energy
distributions obtained with and without that detector appearing
in the trigger pattern,
the reference population being all events where at least
one other detector took part in the trigger.
Both Compton and coincidence techniques give consistent
ionization threshold measurements within 0.2 keV.
For GGA1, the values corresponding to an efficiency
of 50\% are 3.7$\pm$0.2 and  3.5$\pm$0.1 keV,
respectively.
This represents a significant improvement compared to
the performance of GeAl6, where the corresponding values
varied between 5.7 and 11 keV during the run and restricted the
analysis to nuclear recoils above 30 keV.
With a 50\% efficiency reached at 4 keV,
the nuclear recoil selection described below reaches its
full efficiency within less than 1\%
for recoil energies above 20 keV.

\begin{figure}[tbp]
\epsfig{file=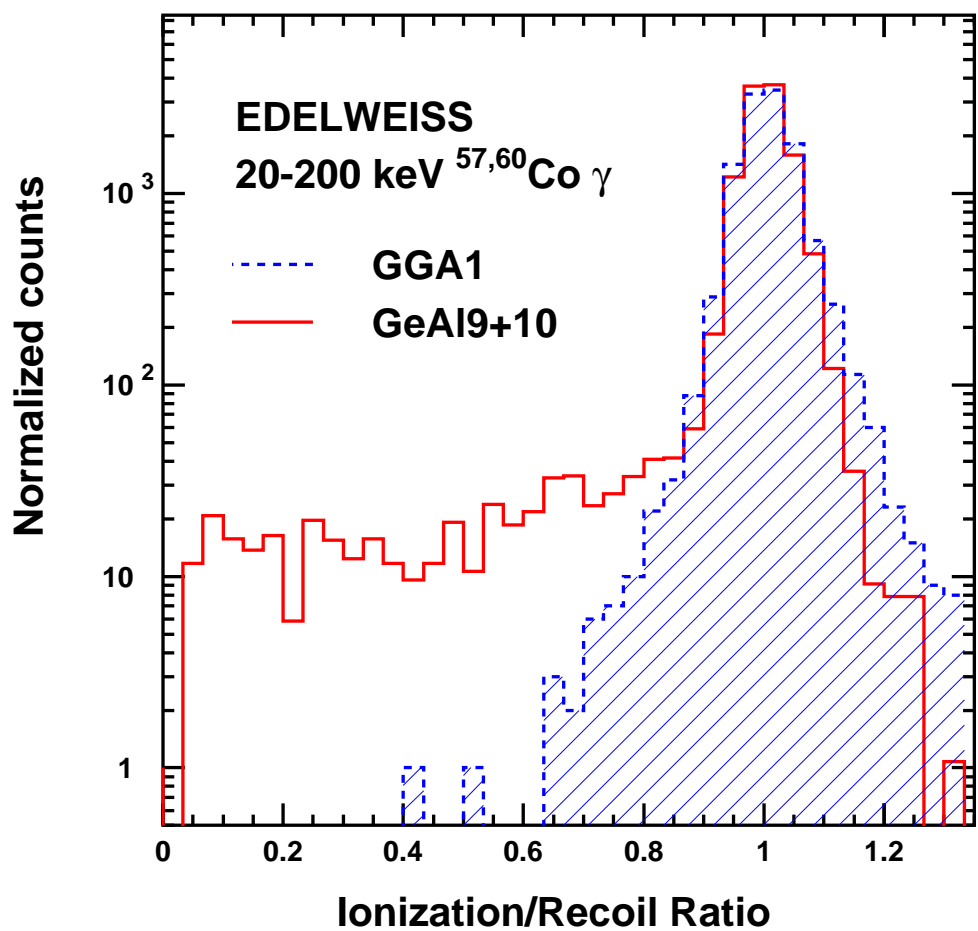
,height=17cm,bbllx=5cm,bblly=10cm,bburx=22cm,bbury=27cm}
\caption[]{Distribution of the ratio of the ionization pulse
height to the recoil energy (quenching factor Q) obtained
by exposing the detectors to $^{57}$Co and $^{60}$Co
$\gamma$-ray sources.
The ratio is normalized to 1 for electron recoils
using the photopeaks of the $^{57}$Co source.
Shaded histogram: detector GGA1.
Line histogram: sum of the detector GeAl9 and GeAl10, normalized
to the same number of entries of GGA1.}
\label{fig-profile} 
\end{figure}

The study of the distribution of the quenching factors $Q$
(the ratio of the ionization signal to the recoil energy,
calculated as in~\cite{bib-edw2000})
recorded in the presence of a $\gamma$-ray source
revealed problems with the charge collection in GeAl9 and GeAl10.
Fig.~\ref{fig-profile} shows the $Q$ distribution for recoils
between 20 and 200 keV recorded with $^{57,60}$Co sources,
for the different detectors.
GeAl9 and GeAl10 have been added together,
since they display a very similar behavior.
The distributions are normalized to the number of entries
in GGA1.
The distribution for GGA1 is centered on 1, as expected
by construction for electron recoils.
The distribution for GeAl9 and GeAl10 show a narrower peak
centered at 1, as expected from the better heat resolution.
However, a relatively flat tail of events with Q values
ranging from 0 to 1 is observed.
While 1.3 and 2.2\% of events have $Q$ values below 0.5
in GeAl9 and GeAl10, this fraction is approximately 0.01\%
in GGA1.
The tail amplitudes do not depend on the recoil energy range.
On the basis of the $^{57,60}$Co calibration runs
it can be expected -- and later experimentally verified --
that these will produce fake nuclear
recoil events (Q $\sim$ 0.3, see below) at a rate of
a few events per kg$\cdot$d in GeAl9 and GeAl10.
It was therefore decided that only the GGA1 data
would be used for deriving a limit on WIMP interactions.

Although the suppression of charge collection problems
in GGA1 may indicate
that the amorphous layer helps prevent them,
as suggested by~\cite{bib-shutt},
more thorough tests are needed before reaching any
firm conclusion.

\mbox{}\\\noindent{\large\bf Fiducial volume and acceptance}\\

As in Ref.~\cite{bib-edw2000}, a fiducial volume
is defined in order to exclude events occurring in the outer
perimeter of the detector as it is
more exposed to external sources of radioactivity
and to charge collection problems.
The selection cut is the same: more than 75\% of the
total charge must be collected on the center electrode.
Here also, two methods are used to measure the
fraction of the total detector volume thus defined.
The first one uses the data collected with
the $^{252}$Cf source: the fraction of nuclear recoil
candidates passing the fiducial cut is compared with the
results of a Monte Carlo simulation of the neutron interactions
in the detector.
The second method exploits the uniformity of the
$^{65}$Zn and $^{68}$Ge decays within the detector volume.
The fraction of the total intensity of the 8.98 and 10.37 keV
peaks selected by the fiducial acceptance is then equal
to the fiducial volume fraction.
The two methods give identical results for GGA1
(57 $\pm$ 3\%) and agree within 3\%
for the other two detectors.

The acceptance for nuclear recoils is defined both in
terms of ranges in $Q$ and recoil energies.
The neutron calibration of the three detectors
confirms the parametrization used in Ref.~\cite{bib-edw2000},
namely,
the center of the band is given by $0.16(E_R)^{0.18}$,
where $E_R$ is the recoil energy in keV,
and its width is equal to that
predicted from the propagation of the heat and ionization
resolutions added in quadrature with a constant {\em rms}
spread of $\sim$0.035.
Again, the width of the band is set to $\pm$1.645$\sigma$.
It was verified on the neutron data that this selection
does correspond to an efficiency of 90\%.

The lower bound of the recoil energy range for the
selection of nuclear recoil is set to 20 keV,
based on the same arguments as in Ref.~\cite{bib-edw2000}:
the efficiency to nuclear recoils should be as uniform
as possible within the band, and it should exclude
regions where the $\gamma$-ray rejection, estimated by propagating
the experimental heat and ionization resolutions, is expected
to be worse than 99.9\%.

Given that a background of events with improper
charge collection has appeared in GeAl9 and GeAl10,
and that it has a flat distribution in both $Q$
and recoil energy,
one could expect a similar behavior, albeit at a lower
level, in GGA1.
Therefore, some care must be taken in the definition of
the upper bound of the recoil energy range of WIMP candidates.
The natural choice is to calculate, using the prescription
of Ref.~\cite{bib-lewin}, the upper bound
corresponding to 95\% of all WIMP-induced
recoils above 20 keV.
Using the standard halo and nuclear form factor
parameters\footnote{
The halo parameters are a
local WIMP density of 0.3 GeV/c$^2$/cm$^3$,
a {\em rms} velocity of 270 km/s,
an escape velocity of 650 km/s
and an Earth-halo relative velocity of 230 km/s.
The Helm parametrization of the nuclear form factor
is used with the recommended values of $a = 0.52$~fm,
$s = 0.9$~fm and $c = 1.23A^{1/3}-0.6$~fm.
See Ref.~\cite{bib-lewin} for details.},
this bound depends on the mass of the WIMP
and varies from 33 keV at 20 GeV/c$^2$ to
86 keV at 100 GeV/c$^2$ and saturates slightly above
110 keV at masses above 10 TeV/c$^2$.

Within the fiducial volume (57\% of 318.5~g),
the acceptance for nuclear recoils
from WIMP interactions thus corresponds to 90\% (width in $Q$)
times 95\% of all recoils above 20 keV
(mass-dependent recoil energy range).

\mbox{}\\\noindent{\large\bf Results and Discussion}\\
%%%%%%%%% Data set

The low-background physics data consists of
all physics runs recorded over a period from
February to May 2002.
The physics data-taking period started a few months after the
installation of the detectors in the Laboratoire Souterrain
de Modane, after a period of optimization of resolution,
calibration and of long exposure to an intense
$^{60}$Co source while all electrodes were shorted.
The running conditions were kept as homogeneous as possible,
until the run was interrupted by an accidental warm-up of
the detector.
In addition to a constant monitoring of the data,
the homogeneity of the running conditions was checked
with $\sim$weekly $^{57}$Co calibrations and two
neutron calibrations.
The total physics run time at low background
corresponds to 54 days, 
of which 2.0\% are lost due to the regular shorting
of the electrodes to prevent the accumulation of
space charge, 3.9\% are lost due to the dead-time
of the data acquisition and 6.2\% are lost in a few
hour-long episodes where the microphonics noise reached unacceptable
levels, as attested by a strong deterioration of the
baseline resolutions.
The total exposure is thus \kgdt\ kg$\cdot$d, of which
\kgdc\ is in the center fiducial volume.
The exposure corrected for the acceptance of the
nuclear recoil band is \kgdeff\ kg$\cdot$d.

%The low-background data of GeAl9 and GeAl10 confirmed
%the prediction of the $^{57,60}$Co runs concerning
%incomplete charge collection.

\begin{figure}[tbp]
\vspace{-1cm}
\epsfig{file=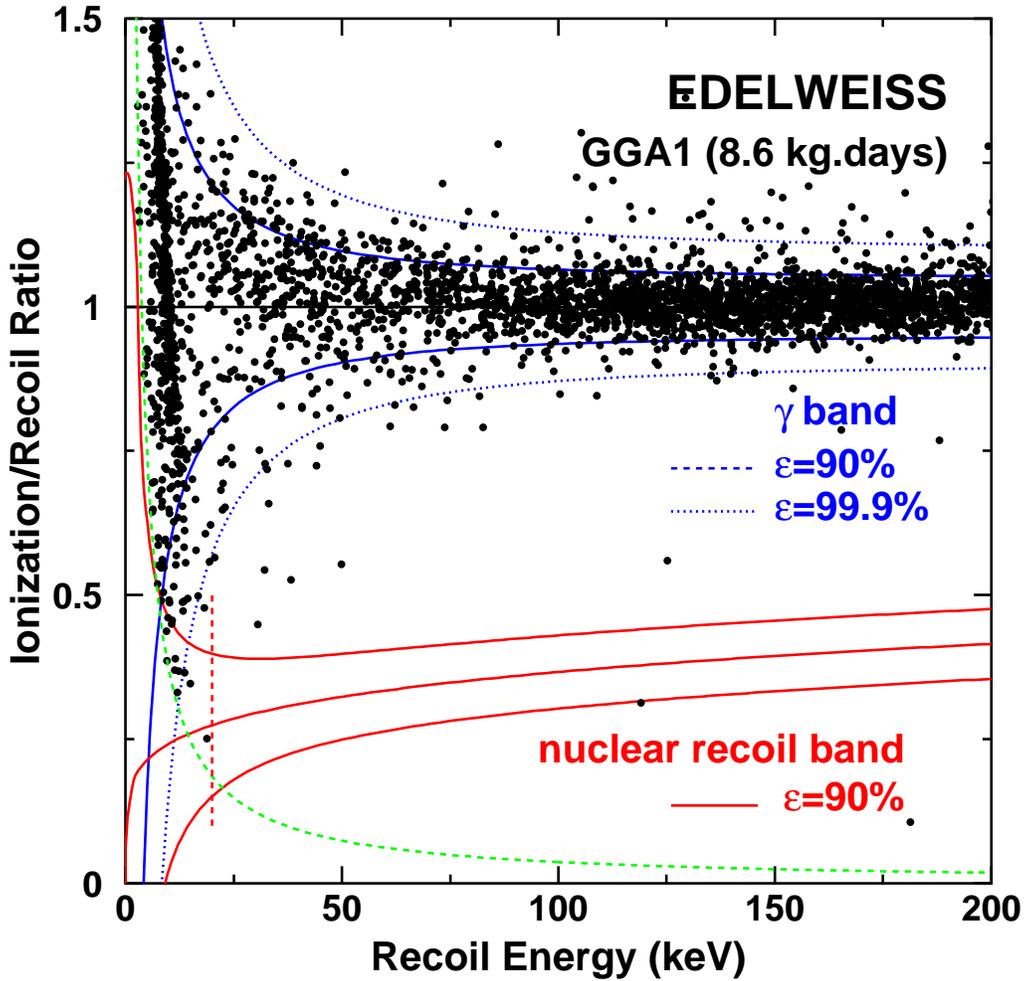
,width=18cm,bbllx=1cm,bblly=5cm,bburx=24cm,bbury=28cm}
\caption[]{Distribution of the quenching factor
(ratio of the ionization signal to the recoil energy)
 as a function of the recoil energy
from the data collected in the center fiducial volume
of the 320~g EDELWEISS detector GGA1.
The exposure of the fiducial volume corresponds to \kgdc\ kg$\cdot$d.
Also plotted as full lines are the $\pm$1.645$\sigma$ bands
(90\% efficiency) for photons and for nuclear recoils.
The 99.9\% efficiency region for photons is also shown (dotted line).
The hyperbolic dashed curve corresponds to 3.5 keV ionization energy
and the vertical dashed line to 20 keV recoil energy.}
\label{fig-fond} 
\end{figure}

The data recorded in the fiducial volume of
GGA1 are shown in Fig.~\ref{fig-fond}.
The ionization-to-recoil energy ratios are plotted
as a function of the recoil energy.
Only events triggered by GGA1 alone and with an ionization
energy above 3.5 keV (hyperbolic dashed line) are shown.
The 99.9\% acceptance band for photons shown as a dotted line
is the result of a simple propagation of
the average heat and ionization resolution,
assuming a Gaussian dispersion.
The population of events around Q$\sim$0.5
associated with low-energy $\beta$ and $\gamma$
surface events,
so prominent in our previous data of Ref.~\cite{bib-edw1997}
and less so in Ref.~\cite{bib-edw2000},
is only represented here by 4 to possibly 5
events.
The correct interpretation of these events would
require a significant increase in exposure, given
the low level of background reached with the
present detector.
The same is true for the interpretation of the
three events below Q=0.7 and with recoil energies
between 119 and 182 keV.

The event at 119 keV and $Q$=0.3 is lying at -1.646$\sigma$
of the centre of the nuclear recoil band.
Given the uncertainty in the experimental determination
of $\sigma$, we conservatively choose to consider
this event as a nuclear recoil,
entering in the acceptance for WIMP masses above
10 TeV/c$^2$.
For lower WIMP masses, no events are in the nuclear
recoil band.

The absence of events in the defined acceptance is
interpreted in terms of upper limits at 90\% CL on
WIMP-nucleon scattering cross sections
for $M_W$ $<$ 10 TeV/c$^2$ following the
prescriptions of Ref.~\cite{bib-lewin} with the
standard halo and nuclear models described above.

\begin{figure}[tbp]
\vspace{-3.0cm}
\epsfig{file=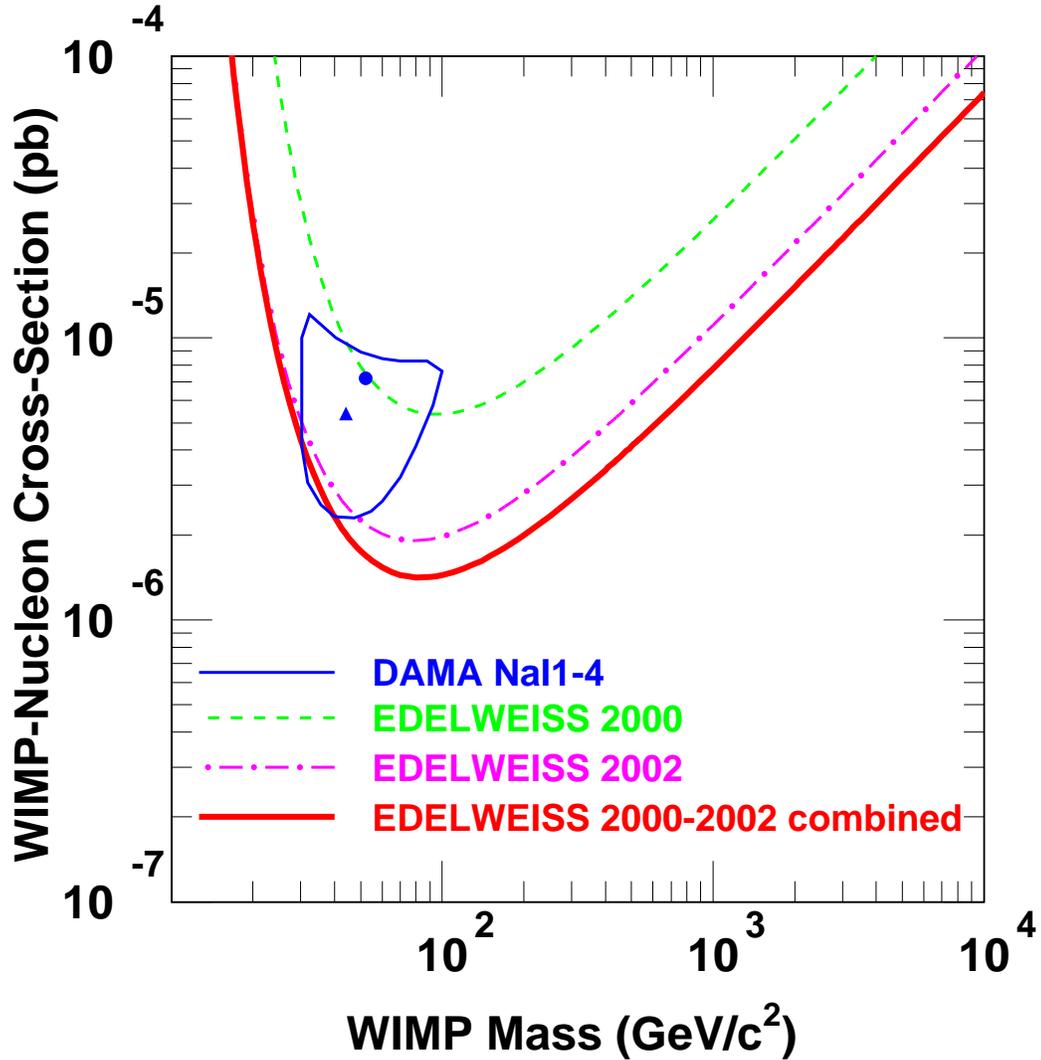
,width=18cm,bbllx=2cm,bblly=6cm,bburx=22cm,bbury=26cm}
\caption[]{
Spin-independent exclusion limits (dark solid curve)
obtained by combining our 2000 data from
Ref.~\protect\cite{bib-edw2000} with the present data,
for a total exposure of 11.7 kg$\cdot$d.
Dashed curve: previous EDELWEISS data~\protect\cite{bib-edw2000} 
re-analyzed using the new definition of the upper bound of the recoil
energy range (acceptance of 95\%).
Dash-dotted curve: present 2002 data.
Closed contour: allowed region at 3$\sigma$ CL from the DAMA1-4
annual modulation data~\protect\cite{bib-dama}.
The full circle and triangle within this contour are
defined in the text.
}
\label{fig-excledw} 
\end{figure}

The limit as a function of WIMP mass
is shown in Fig.~\ref{fig-excledw},
where it is compared with the previous result
obtained with an effective exposure of 4.3 kg$\cdot$d
of the GeAl6 detector~\cite{bib-edw2000}.
The limit resulting from the combination of the
two measurements, corresponding to
an effective exposure of 11.7 kg$\cdot$d is also shown.
The 3$\sigma$ contour corresponding to the annual
modulation effect of DAMA NaI1-4~\cite{bib-dama} is shown:
the black circle marks the central value
of that measurement at $M_W$ = 52 GeV/c$^2$
and $\sigma_n$ = 7.2$\times10^{-6}$~pb.
The present combined results are incompatible with
the interpretation of the modulation effect
in terms of a WIMP behaving
according to the standard phenomenological model
of Ref.~\cite{bib-lewin}.
While \numcin\ nuclear recoils should have been observed
between 20 and 64 keV, none are observed.
The Poisson probability of such a fluctuation
is \proscin\%.
The black triangle on Fig.~\ref{fig-excledw}
at $M_W$=44 GeV/c$^2$ and $\sigma_n$ = 5.4$\times10^{-6}$~pb
corresponds to the most likely value quoted
by DAMA when they combine their modulation results
with their limit achieved using pulse shape discrimination
in NaI~\cite{bib-nai}.
It is also incompatible with the present EDELWEISS results,
the Poisson probability of observing no events from
a prediction of \numqua\ events being \prosqua\%.
The only remaining part of the 3$\sigma$ \mbox{NaI1-4}
DAMA zone corresponds to neutralino masses below the limit
of 45 GeV/c$^2$ obtained at LEP~\cite{bib-acclim}.
Clearly, the standard prescriptions of Ref.~\cite{bib-lewin}
fail at reconciling the EDELWEISS and DAMA
experimental results.

\begin{figure}[tbp]
\vspace{-3.0cm}
\epsfig{file=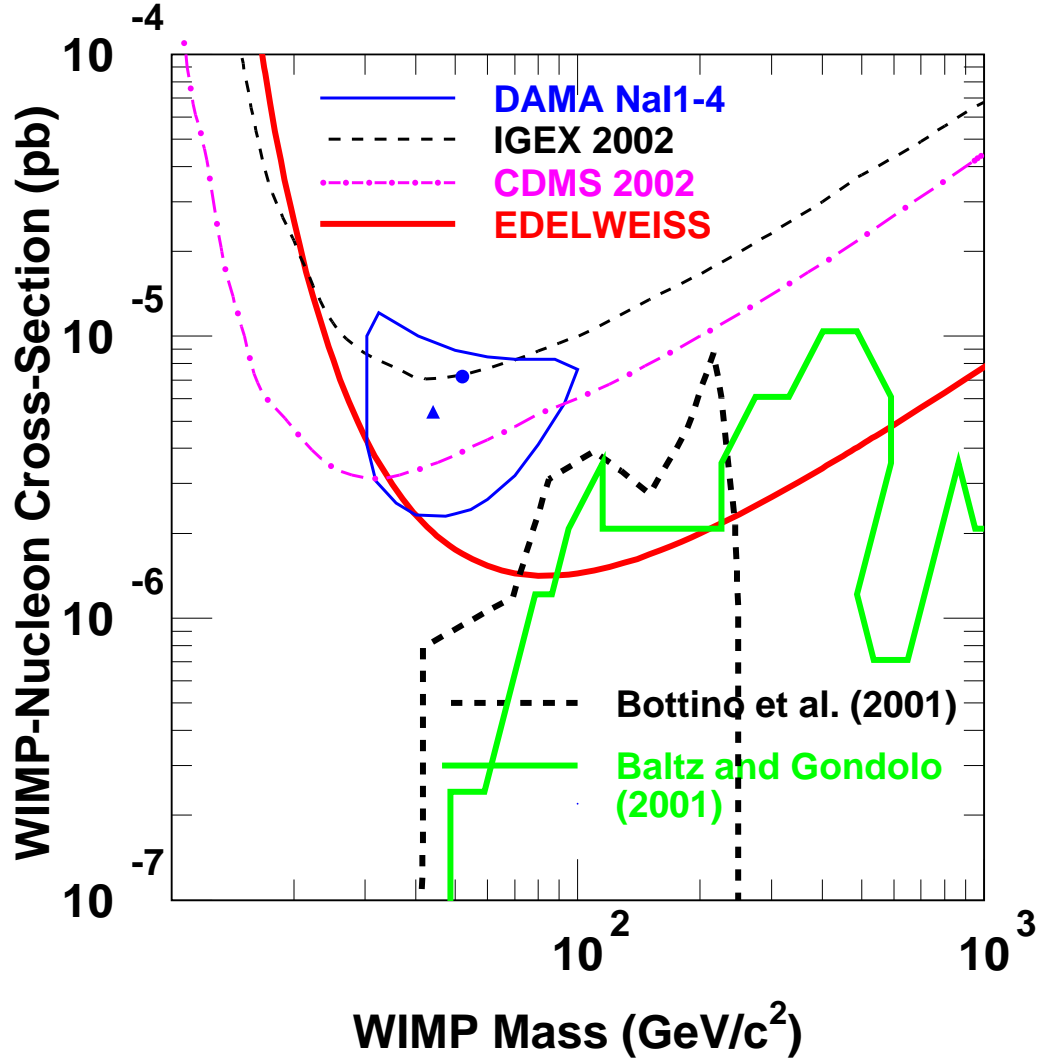
,width=18cm,bbllx=2cm,bblly=6cm,bburx=22cm,bbury=26cm}
\caption[]{
Combined EDELWEISS spin-independent exclusion limits (dark solid curve)
compared with published limits from other experiments and theoretical
calculations.
Dashed curve: Ge diode limit from IGEX~\protect\cite{bib-igex}.
Dash-dotted curve: CDMS limit with statistical subtraction of
the neutron background~\protect\cite{bib-cdms2}.
Closed contour: allowed region at 3$\sigma$ CL from the DAMA1-4 annual
modulation data~\protect\cite{bib-dama}.
Two regions spanned by some of the supersymmetric model calculations
of Refs.~\protect\cite{bib-bottino,bib-gondolo} are also shown.
}
\label{fig-exclus} 
\end{figure}

In Fig.~\ref{fig-exclus} the combined EDELWEISS
limit is compared to those obtained
by other direct WIMP searches.
The EDELWEISS sensitivity for spin-independent
WIMP-nucleon interaction is the best achieved so
far by any dark matter search
for masses above 35 GeV/c$^2$.
Furthermore, the EDELWEISS data start to probe some
of the supersymmetric
models predicting the highest interaction rates.
Fig.~\ref{fig-exclus} shows as an example the range
of masses and cross-sections allowed in the
calculations of Refs.~\cite{bib-bottino,bib-gondolo},
where relaxed conditions of unification at the
GUT scale yield higher upper bounds for $\sigma_n$
(for comparison, see e.g. Ref.~\cite{bib-ellis}).

\mbox{}\\\noindent{\large\bf Conclusion}\\
%%%% conclusion

The EDELWEISS collaboration has searched for nuclear recoils
due to the scattering of WIMP dark matter using a 320~g
heat-and-ionization Ge detector operated in a low-background
environment in the Laboratoire Souterrain de Modane.
After a combination with our previous data~\cite{bib-edw2000},
the achieved sensitivity is so far the best for all direct
searches for WIMP masses above 35 GeV/c$^2$.
The limit obtained on WIMP-nucleon interaction
cross-sections as a function of WIMP mass 
is based on the absence of events
in the recoil energy range relevant for WIMP masses
below 10 TeV/c$^2$
and does not rely on any background subtraction.
The combined EDELWEISS result is incompatible
at more than \probqua\% CL with a WIMP of mass 44 GeV/c$^2$
and a nucleon scattering cross-section of
5.4$\times10^{-6}$pb
reported by the experiment DAMA~\cite{bib-dama}
based on the same standard nuclear physics and
astrophysical assumptions.
Furthermore, the EDELWEISS experiment excludes
a first sample of supersymmetric models
predicting the highest WIMP-nucleon
interaction rates~\cite{bib-bottino,bib-gondolo}.

\section*{Acknowledgments}
The help of the staff of the Laboratoire Souterrain
de Modane and of 
the participating laboratories is gratefully acknowledged.
This work has been partially funded by the EEC Network program under
contract ERBFMRXCT980167.

\newpage

\end{document}